\documentclass[12pt]{iopart}

\usepackage{amsfonts,iopams,cite}

\usepackage[latin1]{inputenc} 
\usepackage[T1]{fontenc}
\usepackage[english]{babel}
\usepackage[pdftex]{graphicx}

\expandafter\let\csname equation*\endcsname\relax
\expandafter\let\csname endequation*\endcsname\relax
\usepackage{amsmath} 							
\usepackage{amssymb}							
\usepackage[colorlinks=false]{hyperref}	
\usepackage{indentfirst}						
\usepackage{epigraph}							
\usepackage{amsthm}	

\usepackage{dcolumn}
\usepackage{color}
\usepackage{epstopdf}

\newcommand{\beq}{\begin{equation}}
\newcommand{\eeq}{\end{equation}}
\renewcommand{\b}[1]{\mathbf{ #1}}															
\newcommand{\de}{\mathrm{d}}																	
\newcommand{\op}[1]{\hat{\mathcal{#1}}}														
\newcommand{\cop}[1]{#1^{\dagger}\!}																
\newcommand{\aop}[1]{#1}														
\newcommand{\up}{\uparrow}														
\newcommand{\down}{\downarrow}												
\newcommand{\m}[1]{\langle #1 \rangle}													
\newcommand{\rom}[1]{\uppercase\expandafter{\romannumeral #1\relax}} 
\renewcommand{\mod}{\,\mathrm{mod}}

\begin{document}

\title[
]{Semimetal-superfluid quantum phase transitions in 2D and 3D lattices with Dirac points}

\author{G. Mazzucchi}
\address{Dipartimento di Fisica, Universit\`a di Trento, via Sommarive 14, I-38123 Povo, Italy.}
\address{SISSA, via Bonomea 265, I-34136 Trieste, Italy.}
\address{Clarendon Laboratory, Department of Physics, University of Oxford, Parks Road, Oxford OX1 3PU, UK.}
\ead{gabriel.mazzucchi@physics.ox.ac.uk}

\author{L. Lepori}
\address{Departament de F\'{i}sica, Universitat Aut\`{o}noma de Barcelona, E-08193 Bellaterra, Spain.}
\address{IPCMS (UMR 7504) and ISIS (UMR 7006), Universit\'{e} de Strasbourg and CNRS, Strasbourg, France.}
\ead{llepori@unistra.fr}

\author{A. Trombettoni}
\address{CNR-IOM DEMOCRITOS Simulation Center, Via Bonomea 265, I-34136 Trieste, Italy.}
\address{SISSA, via Bonomea 265, I-34136 Trieste, Italy.}
\address{INFN, Sezione di Trieste, I-34127 Trieste, Italy.}
\ead{andreatr@sissa.it}

\begin{abstract}
We study the superfluid properties of attractively interacting fermions 
hopping in a family of $2D$ and $3D$ lattices in the presence of 
synthetic gauge fields having $\pi$-flux per plaquette. 
The reason for such a choice is that the $\pi$-flux cubic lattice displays 
Dirac points and that decreasing the hopping coefficient 
in a spatial direction (say, $t_z$)
these Dirac points are unaltered: it is then 
possible to study the $3D$-$2D$ interpolation towards the 
$\pi$-flux square lattice. We also consider 
the lattice configuration providing the continuous interpolation 
between the $2D$ $\pi$-flux square lattice and the honeycomb geometry.
We investigate by a mean field analysis the effects of interaction and dimensionality on 
the superfluid gap, chemical potential and critical temperature, 
showing that these quantities continuously vary 
along the patterns of interpolation.  
In the two-dimensional cases at zero temperature and at half filling 
there is a quantum phase transition occurring at a critical (negative) 
interaction $U_c$ presenting a linear critical exponent for the gap 
as a function of $|U-U_c|$. We show that in three dimensions 
this quantum phase transition is again retrieved, pointing out that 
the critical exponent for the gap changes from $1$ to $1/2$ for each 
finite value of $t_z$.
\end{abstract}

\maketitle

\section{Introduction}

Trapped ultracold atoms provide an ideal system to implement quantum 
simulations of interacting lattice systems~\cite{bloch08}: 
a key tool is given by the use of optical lattices~\cite{lewenstein12}, 
which are used for instance to study Josephson dynamics~\cite{cataliotti01} and 
Mott-superfluid transitions~\cite{greiner02}. The low-energy dynamics 
of ultracold bosons or/and fermions in deep optical lattices 
is described by Bose or/and Fermi Hubbard-like 
models~\cite{jaksch98,hofstetter2002}. In this way, a two-component 
Fermi mixture in a deep optical lattice allows for the physical implementation 
of Fermi-Hubbard physics with ultracold fermions~\cite{esslinger10}, since 
the ratio between the
tunneling rate and the on-site energy can be controlled with 
high precision~\cite{morsch06,bloch08}. 
The dimensionality, the shape of the lattice and the anisotropy of the system 
can be as well tailored: e.g., optical lattices with different strengths 
in each direction can be synthesized.

A further boost for the field of quantum simulations with ultracold atoms 
has been given by the possibility to simulate Abelian and non-Abelian 
gauge fields~\cite{dalibard11}, which are the main subject of the present 
Journal of Physics B Special Issue. The synthesis of gauge fields very much 
enlarges the versatility of the use of ultracold atoms, allowing in perspective 
to explore new types of topological phases and to implement strongly correlated states 
relevant for topological quantum computation~\cite{nayak08}. Synthetic magnetic and electric
fields acting on neutral atoms have already been implemented using 
spatially dependent optical couplings between internal states of the 
atoms~\cite{lin09,lin11}.
This technique has been applied to single-component Bose 
gases~\cite{lin09,lin11,lin09_2} and to Bose-Einstein condensates with
two components~\cite{lin11_2,fu11}. Spin-orbit coupled 
Fermi gases were as well recently realized~\cite{wang12,cheuk12}. 

Artificial gauge potentials can be also studied in the presence of optical 
lattices. As discussed in~\cite{jaksch03}, by 
suitably engineering external laser fields one can control the phase 
accumulated by an atom tunnelling across a plaquette, resulting 
in a non-vanishing artificial magnetic field:
a gauge potential amounts in generally to complex hopping rates on the bonds. Proposals 
for implementing non-Abelian gauge potentials in optical lattices 
were also discussed~\cite{osterloh05}. In the non-Abelian case 
the internal degrees of freedom where the gauge potential is acting on 
are generally hyperfine levels of suitably chosen atoms: e.g, an advantageous choice could be 
provided by earth alkaline atoms as $Yb$~\cite{gerbier10}.  

Experimental results have been recently obtained for the simulation of magnetic fields for 
ultracold atoms in optical lattices. Using Raman-assisted tunneling 
in an optical superlattice, a large tunable effective magnetic fields 
for ultracold atoms was generated and the ground state of the system 
studied~\cite{aidelsburger11}. A one-dimensional lattice with a controllable 
complex tunneling matrix element was obtained from 
a combination of Raman coupling 
and radio-frequency magnetic fields in~\cite{jimenez12}, while in~\cite{hauke12} a method bases 
on a shaken spin-dependent square lattice was used to create strong non-Abelian gauge potentials. 

A promising and challenging application of synthetic gauge fields is to the 
tunable experimental realization of systems 
relevant for high energy physics and quantum gauge 
theories~\cite{zhu,maeda09,bermudez10,cirac10,LMT,kapit11,mazza12,Dyn1,Dyn2,Dyn3,Dyn4,Dyn5}: 
in perspective new developments could permit to study in a controllable experimental 
set-up part of the phase diagrams of gauge theories even strongly coupled, like QCD~\cite{Rajagopal} \cite{Dyn4}
.
Clearly gauge fields are a basic ingredient to realize this task: up to now 
only static gauge fields have been experimentally simulated, but 
there are recent proposals for dynamical fields~\cite{Dyn1,Dyn2,Dyn3,Dyn4,Dyn5}. 

Another central requirement in this direction is the synthesis of relativistic fermionic matter: 
a proposal for this comes with no doubt from graphene, which has been in the last decade the subject 
of an huge amount of attention~\cite{castro}. This system is  a single layer 
of carbon atoms arranged in an honeycomb lattice structure: 
indeed its peculiar geometric configuration leads to an energy spectrum with two bands 
crossing each other linearly in two isolated inequivalent points of the Brillouin zone, 
the so called \emph{Dirac points}. The low-energy excitations around these points 
show indeed an unique dispersion relation mimicking the behavior of massless Dirac fermions and 
this unusual behavior leads to atypical transport and magnetic properties~\cite{castro}. 

In graphene interaction strength and geometry are fixed by the electronic configuration~\cite{castro}: 
it is then interesting to have a physical system in which such properties are easily controllable. 
A solution to this problem comes from ultracold atoms in optical lattices~\cite{bloch05}: 
the honeycomb structure (and related structures exhibiting Dirac points) 
can be efficiently simulated with a proper arrangements of the lasers~\cite{zhu}
and with a great tunability of parameters as hopping and interaction. 
The simulation of $(2+1)$ relativistic Dirac fermions, obtained by 
using ultracold fermions in honeycomb lattices, has been discussed in detail in 
literature~\cite{zhu,guineas,wu08,lee09,boada11,alba11}. The effect of attractive interaction 
for a two-component Fermi mixture in an honeycomb lattice was also 
investigated~\cite{zhao1,kopnin08,mc,gremaud12,tsuchiya12}. At half filling (i.e., 
when the single-particle energy spectrum is filled up to the Dirac points) a critical value 
of the interaction is needed to have a superfluid gap: at this critical point a semimetal-superfluid 
transition takes place~\cite{zhao1}. The experimental realization of a 
Fermi gas in a tunable honeycomb lattice~\cite{tarruell}, where the Dirac points 
can be moved and eventually merged, open the possibility to study graphene physics with tunable 
interaction and geometry in ultracold atom set-ups.

In the light of the obtained progresses and on-going perspectives, a relevant future 
achievement appears to be the synthesis of semimetallic systems hosting in $3D$ Dirac fermions. 
Graphene is not an immediate candidate for $3D$ generalizations: stacking up graphene sheets, 
because of the coupling along the $z$-direction, Dirac cones are destroyed~\cite{Wallace}. 
The problem of the existence of Dirac points in $3D$ lattices has been studied~\cite{Abr,Manes} 
by analyzing the required symmetries required: from the point of view of ultracold atoms, a set-up 
with a low number of hopping connections (possibly, only hoppings between 
nearest neighbours) is required for practical reason. A relatively simple alternative 
is offered by the use of gauge potentials: it is indeed possible to show that 
a square lattice with a constant magnetic field having a $\pi$-flux (half of the elementary flux) 
on each plaquette has an single-particle energy spectrum displaying Dirac points~\cite{AM}. 
This observation has been translated in proposals to realize massless $(2+1)$ Dirac fermions using 
external gauge proposals, including ultracold 
fermions on a square lattice coupled with properly chosen Rabi fields \cite{hou09}, 
interacting bosons in a $2D$ lattice produced 
by a bichromatic light-shift potential with an additional effective magnetic field~\cite{lim08} 
and bosons with internal energy levels in a tripod configuration~\cite{Juzeliunas}.

The $2D$ set-up of a square lattice with $\pi$-flux can be generalized to three dimensions~\cite{Has,Zou} 
considering a cubic lattice with $\pi$-flux per plaquette, obtained via a magnetic field oriented 
along the diagonal of the cube: this scheme was the starting point for work about 
chiral spin liquids~\cite{Zou} and topological insulators and superconductors \cite{Hosur}. 
The energy spectrum still shows Dirac points and 
$3D$ Dirac fermions~\cite{LMT}. Therefore using a two-component Fermi gas 
in a constant magnetic field acting on both the components one can have a semimetallic behaviour 
at half filling: adding an attractive interaction one obtains a $3D$ version of the semimetal-superfluid 
transition studied in $2D$~\cite{zhao1,kopnin08,mc,gremaud12,tsuchiya12}.

The present paper aims to analyze at a mean field level the superfluidity in a family of 
$2D$ and $3D$ lattice systems with Dirac points and in 
the presence of an attractive point-like interactions 
between the two species of atoms. Since the Dirac points are unaltered by varying the hopping coefficient 
in a spatial direction (say, $t_z$) it is possible to study in this system 
the $3D$-$2D$ interpolation between the $\pi$-flux cubic lattice the 
$\pi$-flux square lattice. We also consider 
the lattice configuration providing the continuous interpolation 
between the $2D$ $\pi$-flux square lattice and the honeycomb geometry. 
Along these patterns of interpolations the mean field equations are solved, studying the behaviour 
of the superfluid gap, the critical temperature and the chemical potential.

\section{The models}

In this Section we introduce the attractive Hubbard model and the lattices 
studied in the rest of the paper; we also briefly 
summarize the mean field treatment, writing down the equations 
for the gap and the number of particles studied in the following Sections. 
The attractive Hubbard Hamiltonian is written as 
\begin{align}\label{fh-ham1}
	\op{H} = -\sum_{\langle i,j \rangle, \sigma} \left( t_{ij} \cop{c}_{i \sigma} 
\aop{c}_{j \sigma} + h.c. \right) - U \sum_i \cop{c}_{i \up} \cop{c}_{i \down} \aop{c}_{i \down} \aop{c}_{i \up}.
\end{align}
In eq. (\ref{fh-ham1}) $\aop{c}_{j \sigma}$ is the fermionic operator destroying a particle in the site 
$i$ of the lattice with (pseudo)spin $\sigma=\up, \down$; the sum 
on the first term in eq. (\ref{fh-ham1}) is on distinct pairs of nearest neighbours and 
$U$ will be assumed to be positive corresponding to 
on-site attractions. The total number of particles is denoted by $N$, and $n$ is the filling 
(number of particles per lattice site): denoting the number of lattice sites by $\Omega$, it is 
$n=N/\Omega$. Within the grand canonical ensemble it is necessary to 
introduce the chemical potential $\mu$ and consider the operator $\op{H}- \mu \op{N}$ for 
describing the system. 

The hopping rates $t_{ij}$ entering eq. (\ref{fh-ham1}) 
are in general complex: in presence of a synthetic field 
$\b{B} = \nabla \times \b{A}$ acting on both the Fermi species, through the Peierls substitution 
one introduces complex hopping rates in the kinetic term of the Hubbard Hamiltonian. 
We will assume that the synthetic gauge potential $\b{A}$ is acting on both species in the same 
way, corresponding to a $U(1) \times U(1)$ gauge potential. 
In particular the hopping rate from the site $i$ to the site $j$ reads 
\begin{align}
t_{ij}=|t_{ij}|\exp \left(- i \int_i^j \de \b{l} \cdot \b{A} \right).
\label{Peierls}
\end{align}
We consider in Section \ref{2sect} a magnetic field along $\hat{z}$, while in Section \ref{3sect} 
the magnetic field is taken along the direction $(1,1,1)$. In Section \ref{3sect} we also 
consider the possibility that $|t_{ij}|$ depends on the direction, having $|t_{ij}| \equiv t$ along 
the directions $\hat{x},\hat{y}$ and $|t_{ij}| \equiv t_z$ along $\hat{z}$: this can be obtained by using 
optical lattices of a different strength in the $\hat{z}$ directions~\cite{iazzi12}.

We denote by $\epsilon(\b{k})$ the single-particle energy spectrum 
of Hamiltonian (\ref{fh-ham1}) with $U=0$, denoted by $\op{H}_0$:
\begin{align}\label{fh-ham1-kin}
	\op{H}_0 = -\sum_{\langle i,j \rangle, \sigma} \left( t_{ij} \cop{c}_{i \sigma} 
\aop{c}_{j \sigma} + h.c. \right).
\end{align}
For a $D$-dimensional cubic lattice (lattice spacing 
is taken equal to $1$) and with $\b{A} = 0$ and $|t_{ij}| \equiv t$ it is e.g.:
$\epsilon(\b{k})=-2t \sum_{\alpha=1}^D \cos{k_\alpha}$.  The single-particle spectrum and the corresponding 
Brillouin zones (BZ) of the $\pi$-flux square and cubic lattices will be discusses respectively 
in Sections \ref{2sect} and \ref{3sect}.

In the Hartree-Fock approximation the interaction term 
$\cop{c}_{i \up} \cop{c}_{i \down} \aop{c}_{i \down} \aop{c}_{i \up}$ is 
replaced by the two particles operator
\begin{equation}
\m{\cop{c}_{i \up} \cop{c}_{i \down}}\aop{c}_{i \down} \aop{c}_{i \up} + \m{\aop{c}_{i \down} \aop{c}_{i \up}}\cop{c}_{i \up} \cop{c}_{i \down} + \m{\cop{c}_{i \up} \aop{c}_{i \up}}\cop{c}_{i \down} \aop{c}_{i \down} +
  \m{\cop{c}_{i \down} \aop{c}_{i \down}}\cop{c}_{i \up} \aop{c}_{i \up}-  \m{\cop{c}_{i \up} \aop{c}_{i \down}}\cop{c}_{i \down} \aop{c}_{i \up} -  \m{\cop{c}_{i \down} \aop{c}_{i \up}}\cop{c}_{i \up} \aop{c}_{i \down},
\label{sost}
\end{equation}
where the $\m{\dots}$ is the grand canonical average computed with the 
(non interacting) Hartree-Fock Hamiltonians~\cite{Annett}. Thanks to translational invariance we can drop 
the $i$-dependence on the mean values in eq. (\ref{sost}); moreover by spin-rotational symmetry 
it is $\m{\cop{c}_{i \down} \aop{c}_{i \up}}=\m{\cop{c}_{i \up} \aop{c}_{i \down}}=0$, 
$\frac{n}{2}=\m{\cop{c}_{i \up} \aop{c}_{i \up}} = \m{\cop{c}_{i \down} \aop{c}_{i \down}}$. The gap 
parameter is as usual defined by $U \m{\aop{c}_{i \down} \aop{c}_{i \up}}=\Delta$~\cite{Annett}. The Hartree-Fock Hamiltonian is then given by 
\begin{align}\label{fh-hf}
	\op{H}_{HF}- \mu \op{N} = &-\sum_{\langle i,j \rangle, \sigma} \left(t_{ij} \cop{c}_{i \sigma} \aop{c}_{j \sigma} + 
h.c. \right) - \tilde{\mu} \sum_{i,\sigma} \cop{c}_{i \sigma} \aop{c}_{i \sigma} + \nonumber \\
	& \quad - \Delta \sum_i \left( 
\cop{c}_{i \up} \cop{c}_{i \down} + \aop{c}_{i \down} \aop{c}_{i \up}\right),
\end{align}
where we defined $\tilde{\mu}= \mu - Un/2$. Moving to momentum space one gets:
\begin{align}\label{fh-hf-fourier}
	\op{H}_{HF}- \mu \op{N} = \sum_{\b{k}\in \mathrm{BZ}, \sigma} \left( \epsilon(\b{k}) - \tilde{\mu}\right) \cop{c}_{\b{k} \sigma} \aop{c}_{\b{k} \sigma} 
- \Delta \sum_{\b{k}\in \mathrm{BZ}} \left( 
\cop{c}_{\b{k} \up} \cop{c}_{\b{k} \down} + \aop{c}_{\b{k} \down} \aop{c}_{\b{k} \up} \right).
\end{align}
Diagonalizing eq. (\ref{fh-hf-fourier})  
and minimizing the free energy ~\cite{Annett} one gets at temperature $T$
\begin{align}\label{mf_eq}
	\left\{ \begin{array}{l}
	\displaystyle \frac{1}{U} = \frac{1}{2 \Omega} \sum_{\b{k}\in \mathrm{BZ}} \frac{1}{E(\b{k})} \tanh\left( \frac{\beta E(\b{k})}{2}\right) \\[7mm]
	\displaystyle n = \frac{1}{\Omega} \sum_{\b{k}\in \mathrm{BZ}} \left[ 1 - \frac{\epsilon_0(\b{k})}{E(\b{k})} \tanh\left( \frac{\beta E(\b{k})}{2}\right) \right]
	\end{array} \right.
\end{align}
where $\beta=1/k_B T$, 
$\epsilon_0(\b{k}) = \epsilon(\b{k}) - \tilde{\mu}$ and \begin{align}
	E(\b{k})=\sqrt{\epsilon_0^2(\b{k}) + \Delta^2}.
\end{align}
is the 
the excitation spectrum.

The solutions of eqs. (\ref{mf_eq}) reveals in general that increasing 
the strength of the attractive interaction $U$, a BCS-BEC crossover takes 
place: the chemical potential $\mu$ 
decreases with $U$ increasing ~\cite{chen05,giorgini08,zwerger12}. Studies of the BCS-BEC crossover for the 
attractive Hubbard model in cubic lattices are available in 
literature~\cite{Singer1996,Sewer2002,Toschi2005,Burovski2006,Chien2008,iazzi12}. 
Notice that a qualitatively correct estimation of the critical temperature (at which 
$\Delta=0$) requires - far from the BCS limit - to use gaussian fluctuation around the mean field 
saddle point~\cite{sademelo93}.  

Solutions of the mean field eqs. (\ref{mf_eq}) on various $2D$ and $3D$ 
lattices with Dirac points will be presented in the 
following Sections.

\section{$\pi$-flux square lattice model}
\label{2sect}
 
In this Section we study the solution of the mean field eqs. (\ref{mf_eq}) for a $2D$ square lattice 
in the presence of a magnetic flux generating a $\pi$-flux per plaquette: this model exhibits 
Dirac points at half filling. 
We also consider a continuous interpolation between the $\pi$-flux square lattice model and the 
honeycomb lattice, on which the attractive Hubbard model has been extensively 
studied~\cite{zhao1,kopnin08,mc,gremaud12,tsuchiya12}.

\begin{figure}[t!]
\centering
\includegraphics[width=0.46\textwidth]{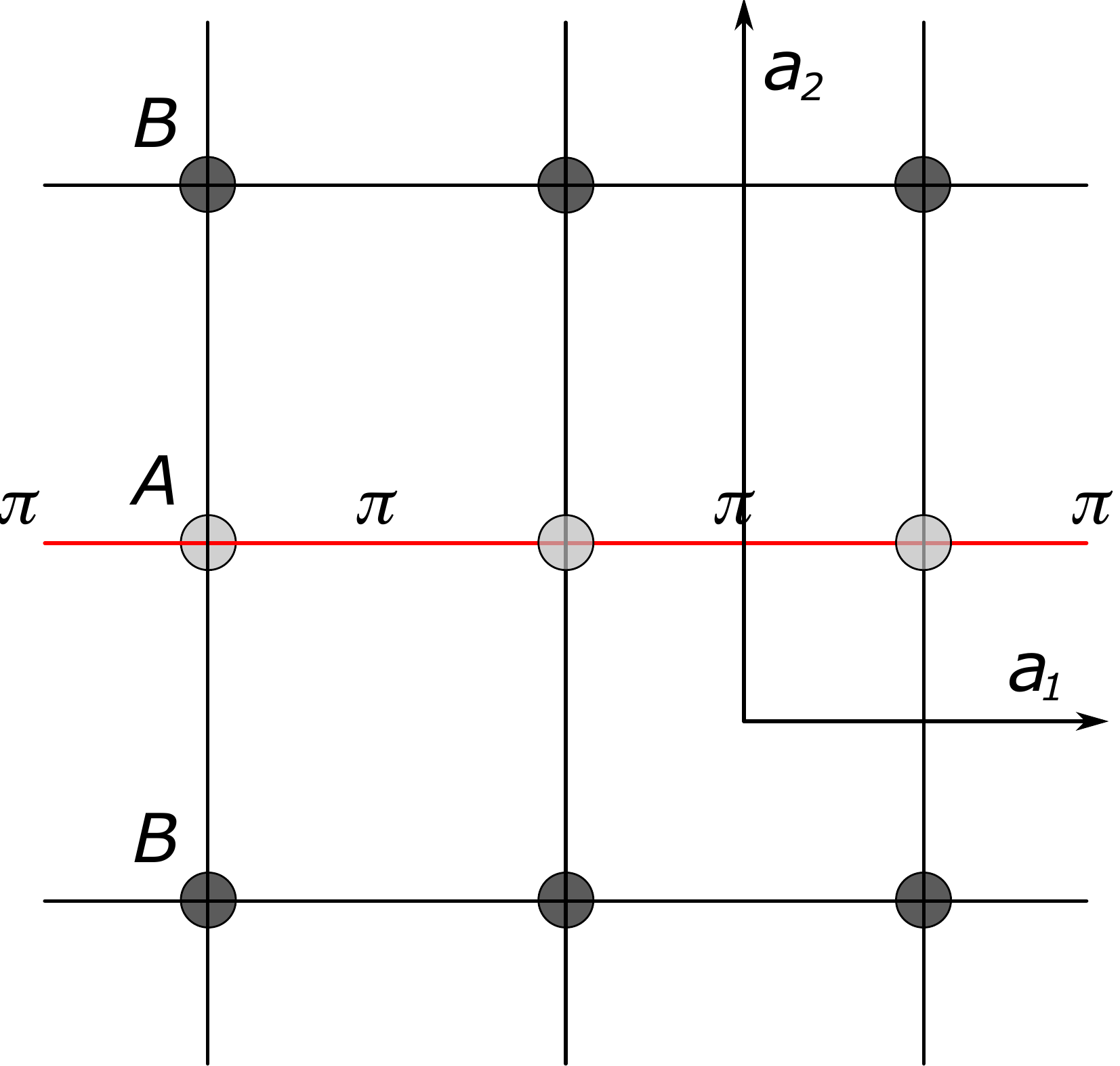}
\qquad
\includegraphics[width=0.46\textwidth]{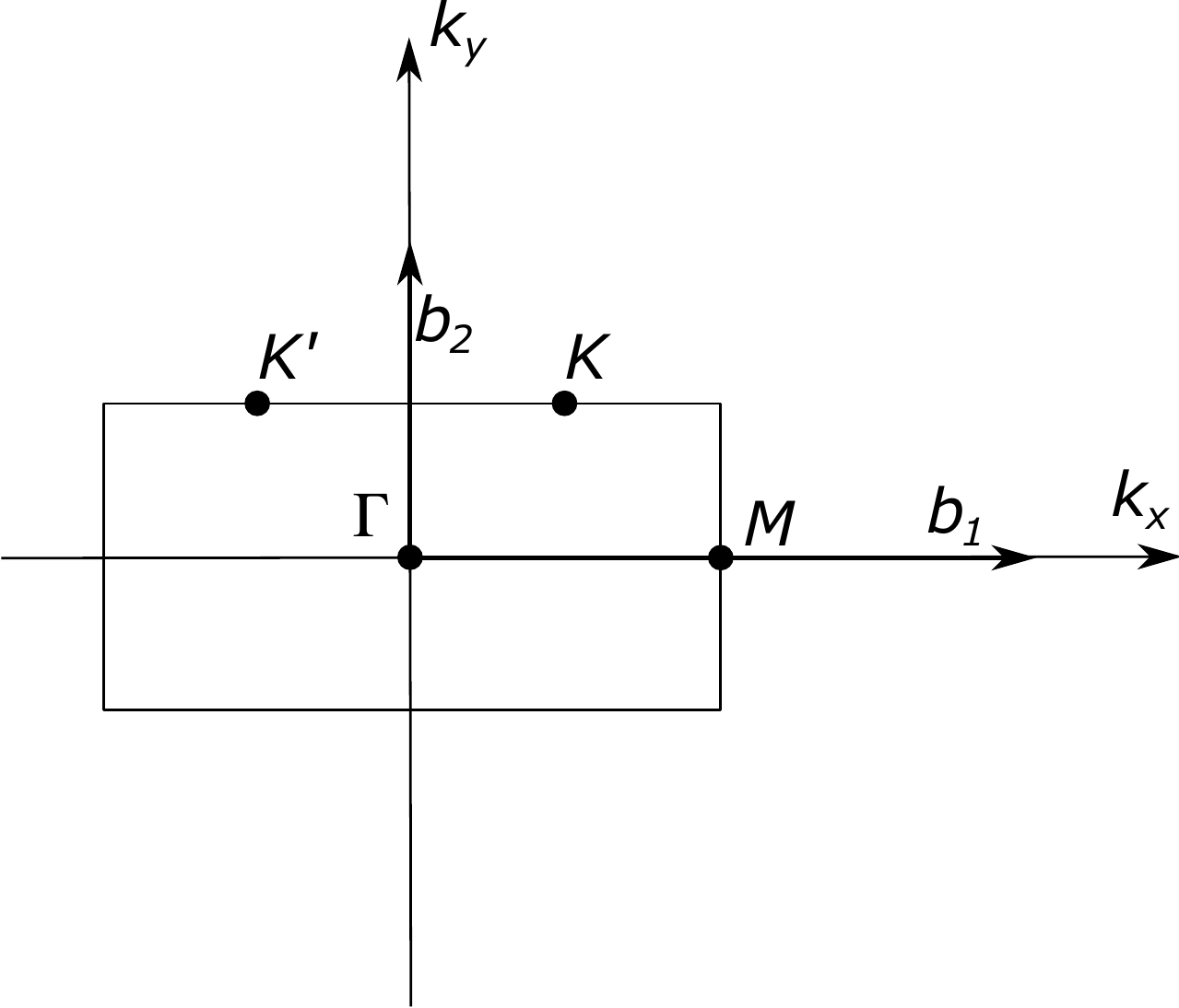}
\caption{$\pi$-flux square lattice model (left) and corresponding first Brillouin zone (right). 
Direct and reciprocal lattice vectors are shown. Hopping amplitudes on the central 
red line on the left figure are complex with phase $\pi$, the hopping amplitudes on black bonds have instead a real phase equal to 1.}
\label{fig:am_lattice}
\end{figure}

A $2D$ lattice model having Dirac cones in the energy dispersion was discussed in~\cite{AM}: 
this model is formulated on a square lattice with 
nearest-neighbour hoppings where an orthogonal magnetic field $\b{B}$ 
is applied such to have on each plaquette a flux half of the fundamental one $\Phi_0$ (from now on we set
$\Phi_0 \equiv 1$). 
Using Peierls substitution one can implement a magnetic field introducing complex hopping amplitudes 
according eq. (\ref{Peierls}), where 
\begin{equation}
\b{A} = \pi \, (-y,0,0)
\label{gland}
\end{equation} 
is the vector potential giving raise to $\b{B}$. The scheme of the phases acquired by a particle 
in a single hopping and with this gauge choice is plotted in Fig. (\ref{fig:am_lattice}). 
These quantities are not gauge invariant, while the total phase acquired around a loop is. 
Notice the doubling of the elementary cell in one space direction and the 
consequent halving of a reciprocal vector.
Indeed a choice for the first magnetic Brillouin zone is that the lattice vectors are:
\begin{align}
	\b{a}_1= d \left( 1 ,0 \right) \quad \mbox{and} \quad \b{a}_2=2 d \left( 0 ,1\right) \, ,
\end{align}
where $d$ is the lattice constant, with the  reciprocal lattice vectors given by
\begin{align}
	\b{b}_1=\frac{2 \pi}{d} \left( 1 ,0 \right) \quad \mbox{and} \quad \b{b}_2=\frac{\pi}{d}\left(0 ,1 \right).
\end{align}
From now on we will set $d = 1$.
Notice that the primitive cell is half with respect to the square lattice with no magnetic flux 
because in presence of $\b{B}$ the translations along $\hat{x}$ and $\hat{y}$ do not commute 
each other any longer and one has to double one of them.

\begin{figure}[t!]
\centering
\vspace{0.5cm}
\includegraphics[width=0.80\textwidth]{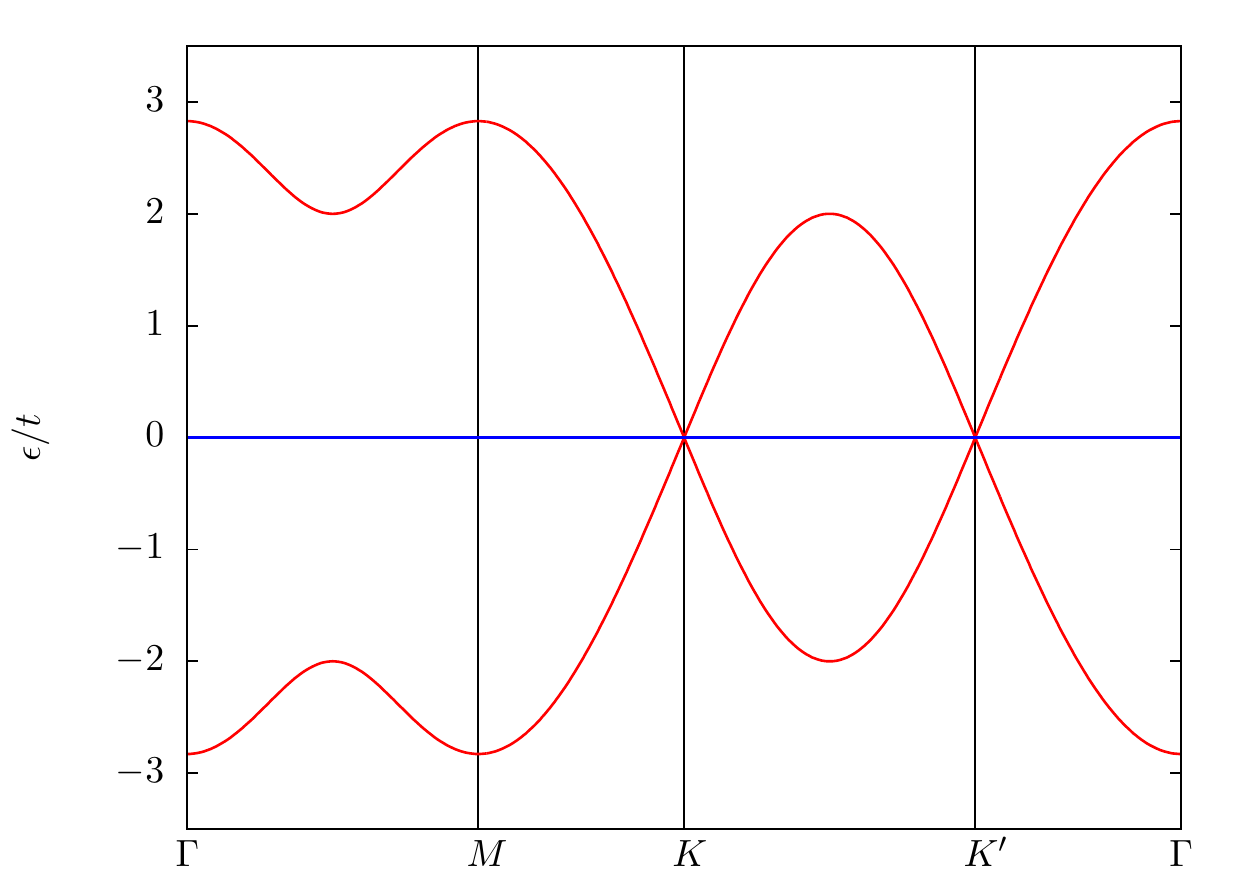}
\caption{Dispersion relation for the tight-binding Hamiltonian $\op{H}_0$ 
on the $\pi$-flux square lattice model. The horizontal line represents the Fermi level at half filling.}
\label{fig:am_bands}
\end{figure}

\begin{figure}[t!]
\centering
\vspace{1.5cm}
\includegraphics[width=0.80\textwidth]{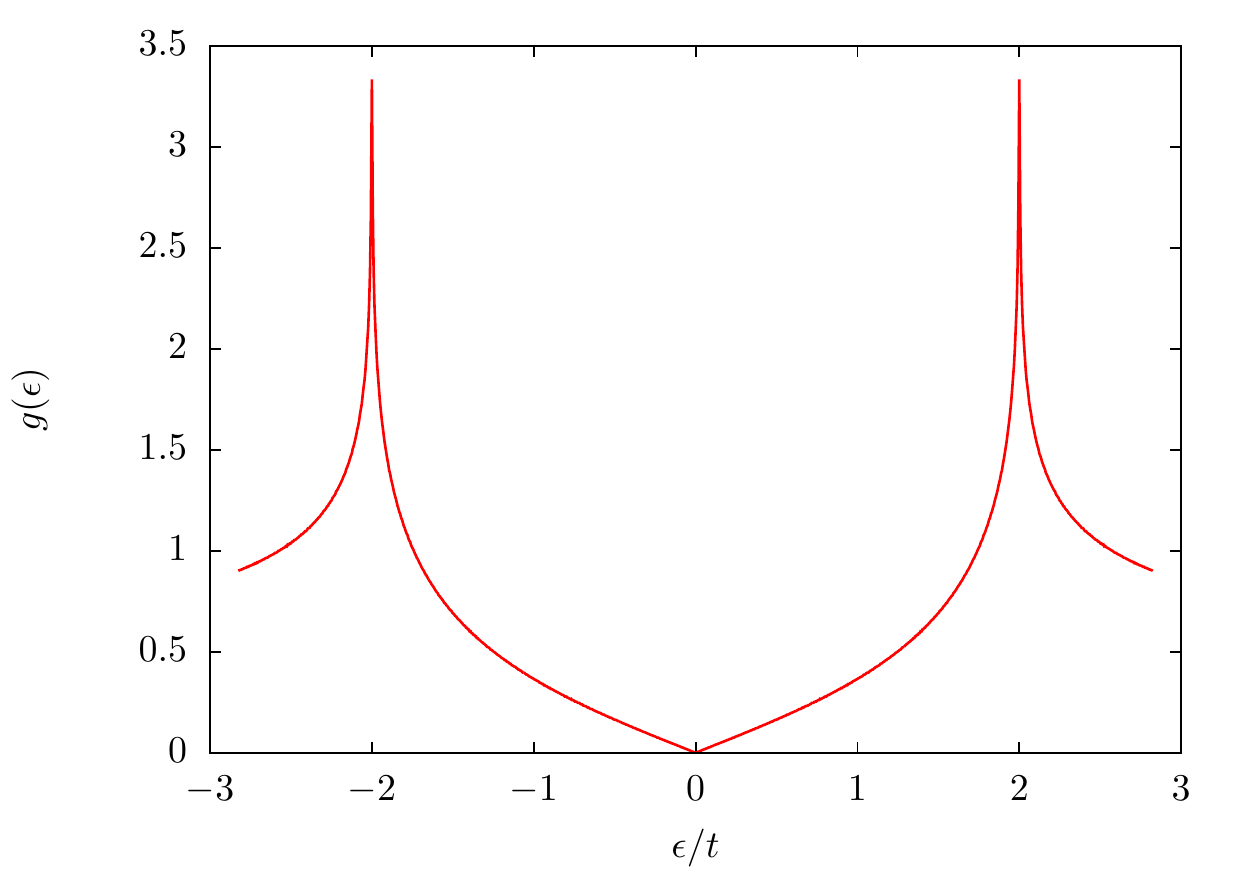}%
\caption{Density of states (normalized to $2$) for the $\pi$-flux square lattice model.}
\label{fig:am_dos}
\end{figure}

The kinetic Hamiltonian $\op{H}_0$, defined by eq. (\ref{fh-ham1-kin}), reads in momentum space:
\begin{align}\label{am_nn}
	\op{H}_0=-2t \sum_{\b{k}\in \mathrm{BZ},\sigma} \begin{pmatrix} \cop{c}_{\b{k} \sigma,a} \cop{c}_{\b{k} \sigma,b}\end{pmatrix}   \begin{pmatrix} \cos k_x & -\cos k_y\\ -\cos k_y  & - \cos k_x \end{pmatrix}\begin{pmatrix} \aop{c}_{\b{k} \sigma,a} \\ \aop{c}_{\b{k} \sigma,b}\end{pmatrix}
\end{align}
where the subscript $a$ or $b$ is related to the $A$ or $B$ sublattices, a sublattice being 
defined by the set of lattice sites having a given set of hopping phases along the bond starting from them. 
The Hamiltonian matrix can be easily diagonalized leading to the single particle spectrum
\begin{align}\label{am_bands}
	\epsilon_{\pm}(\b{k}) = \pm 2  t \sqrt{\cos^2 k_x + \cos^2 k_y} \, .
\end{align}
The band structure and the density of states (normalized to $2$) are shown in Figs. 
(\ref{fig:am_bands}) and (\ref{fig:am_dos}). 
The main property of (\ref{am_bands}) is the presence of two inequivalent Dirac points in 
\begin{align}
	\b{K}=\frac{\pi}{2} \left( 1 ,1 \right) \quad \mbox{and} \quad \b{K}^{\prime}=\frac{\pi}{2} \left( -1 ,1 \right),
\end{align}
with the other two Dirac points at $\frac{\pi}{2} \left( 1 ,1 \right)$ and $\frac{\pi}{2} \left( 1 ,1 \right)$ 
being related to the previous ones by translations of the primary vectors of the reciprocal lattice. 
The dispersion relation near these points is linear:
\begin{align}
	\epsilon_{\pm}(\b{K}+\b{q}) = \pm 2 t \left| \b{q} \right| + \mathcal{O}\left( \left| \b{q} \right| ^3 \right) 
\end{align}
(unlike the honeycomb lattice, there are no quadratic terms in the expansion near the Dirac points). 
The ``speed of light'' is given by $2t$ and the behavior of the density of states near $\epsilon=0$ is:
\begin{align}
	g(\epsilon) \approx  \frac{1}{2 \pi t^2 } \left| \epsilon \right|.
\end{align} 

\begin{figure}[h!]
\centering
\includegraphics[width=0.70\textwidth]{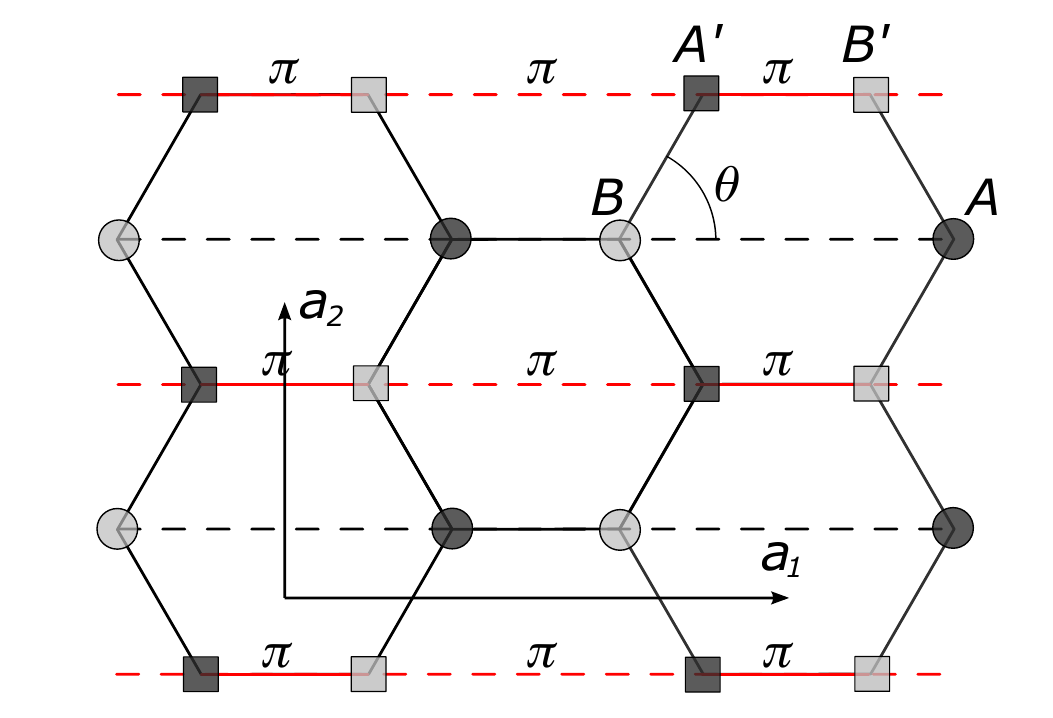}%
\caption{Anisotropic honeycomb lattice with magnetic field: this model allows to interpolate between 
the honeycomb lattice and the $\pi$-flux square lattice model. 
Primitive lattice vectors are shown for $\theta=\pi/3$. Hopping amplitudes on 
bonds
on top, central and bottom 
 red lines have phase $\pi$.}
\label{fig:am_honey_lattice}
\end{figure}

The $\pi$-flux model admits a smooth interpolation with the honeycomb lattice model:
notably along this interpolating path the Dirac points are always present. 
The path that we considered is realized using an anisotropic honeycomb lattice as in 
Fig. (\ref{fig:am_honey_lattice}).  A magnetic field with flux $\pi$ on each trapezoidal plaquette, being the half of the hexagons is added; 
on each hexagon the flux instead vanishes. Notice that the eigenvalues of the Hamiltonian depend only on the total magnetic flux on a single plaquette \cite{Haldane}.

The kinetic part $\op{H}_0$ of the Hubbard Hamiltonian 
is formed by two types of terms: the usual nearest neighbours hopping term $t$ (along solid lines) and 
an anisotropic next to nearest neighbours hopping term (along dashed lines) having $|t_{ij}| \equiv a \, t$. 

Changing the value of the anisotropy parameter $a$ one can pass from the honeycomb lattice 
($a=0$) to a $\pi$-flux square lattice model ($a=1$). 
Notice that the model with $a=1$ is not geometrically 
equivalent to the $\pi$-flux square lattice: 
in order to get exactly the square lattice with $\pi$-flux  
it is necessary to introduce a continuous transformation on the angle $\theta$ 
at the larger basis of the trapezoids, keeping constant the flux on each trapezoid. 
The correct geometry for the $\pi$-flux square lattice model can be recovered with
$\theta=\pi/2$, while the honeycomb lattice is defined by 
setting $\theta=\pi/3$. Notice however that the energy eigenvalues will depend only on $a$ and not on $\theta$. 

Choosing the distance between nearest neighbours to be unitary, the lattice vectors are given by
\begin{align}\label{honey_am_vec}
	\b{a}_1=  \left( 1+ \cos \theta ,0 \right)  \quad \mbox{and} \quad \b{a}_2= \left( 0 ,2 \sin \theta \right)
\end{align}
and the reciprocal lattice vectors are
\begin{align}
	 \b{b}_1=\frac{2 \,  \pi}{(1+  \cos \theta)} \left( 1 ,0 \right)  \quad \mbox{and} \quad \b{b}_2=\frac{\pi}{\sin \theta}\left(0 ,1 \right).
	 \label{honey_am_rec}
\end{align}
Defining 
\begin{align}
	{\cal C}=\cos \theta \qquad \mbox{and} \qquad {\cal S} = \sin \theta
\end{align}
it is possible to write the Hamiltonian $\op{H}_0$ in the Fourier space as
\begin{align}\label{inter_nn}
	\op{H}=-t \sum_{\b{k}\in \mathrm{BZ},\sigma} \begin{pmatrix} \cop{c}_{\b{k} \sigma,a^{\prime}}  \cop{c}_{\b{k} \sigma,a} \cop{c}_{\b{k} \sigma,b^{\prime}} \cop{c}_{\b{k} \sigma,b}\end{pmatrix}  \mathcal{M}   
\begin{pmatrix} \aop{c}_{\b{k} \sigma,a^{\prime}}  \\ \aop{c}_{\b{k} \sigma,a} \\ \aop{c}_{\b{k} \sigma,b^{\prime}} \\ \aop{c}_{\b{k} \sigma,b}\end{pmatrix}
\end{align}
where the subscripts $a$, $a^{\prime}$, $b$ and $b^{\prime}$ are related to the sublattices $A$, $A^{\prime}$, $B$ and $B^{\prime}$ respectively and the matrix $\mathcal{M}$ is given by
\begin{align}
	\mathcal{M}= \begin{pmatrix} {\bf 0} & \mathcal{A} \\ \mathcal{A}^{\dagger} & {\bf 0} \end{pmatrix}.
\label{emme}
\end{align}
In eq. (\ref{emme}) ${\bf 0}$ is the $2 \times 2$ zero matrix and the matrix $\mathcal{A}$ is given by
\begin{align}
	\mathcal{A} = \begin{pmatrix} -\mathrm{e}^{-i k_x}-a \mathrm{e}^{i k_x (1+2{\cal C})} & 
\mathrm{e}^{i ({\cal C} k_x -{\cal S} k_y)}+\mathrm{e}^{i ({\cal C} k_x+{\cal S} k_y)}\\ 
\mathrm{e}^{i ({\cal C} k_x-{\cal S} k_y)}+
\mathrm{e}^{i ({\cal C} k_x+{\cal S} k_y)} & \mathrm{e}^{-i k_x}+a \mathrm{e}^{i k_x (1+2 {\cal C})} \end{pmatrix}.
\end{align}
Diagonalizing $\mathcal{M}$ one gets the energy spectrum
\begin{align}
	&\epsilon_{1,\pm}(\b{k};a) = \pm t \, \sqrt{3 + a^2 + 2 a \cos\left [2 k_x (1+{\cal C})\right] + 
2 \cos \left( 2 k_y \, {\cal S} \right) - 4 (a-1)\cos (k_y \, {\cal S}) \sin \left[ k_x(1+{\cal C})\right] }
\label{interpol_2D_1} \\
	&\epsilon_{2,\pm}(\b{k};a) = \pm t \, \sqrt{3 + a^2 + 2 a \cos\left [2 k_x (1+{\cal C})\right] + 
2 \cos \left( 2 k_y \, {\cal S} \right) + 4 (a-1)\cos (k_y \, {\cal S}) \sin \left[ k_x(1+{\cal C})\right] }.
\label{interpol_2D_2}
\end{align}
Notice that eqs. (\ref{interpol_2D_1}) and (\ref{interpol_2D_2}), using eqs. (\ref{honey_am_vec}) and (\ref{honey_am_rec}), do not depend explicitly on $\theta$.
The results for the $\pi$-flux square lattice model can be recovered imposing 
 $a=1$, while the honeycomb lattice is defined by 
setting $a=0$. In these two limits the energy bands are doubly degenerate. In particular for $a=0$, $\epsilon_{1,\pm}(\b{k};0)$ and $\epsilon_{2,\pm}(\b{k};0)$ are related by the momentum shift $k_x \to k_x +  \pi$ or $k_y \to k_y +  \pi$. 
This is due to the fact that the cell defined by the vectors in (\ref{honey_am_vec}) 
is not a primitive one: 
selecting correctly the primitive cell this fictitious degeneration 
is removed and one gets two non degenerate 
bands. One then finds the spectrum (\ref{am_bands}) for the $\pi$-flux square 
lattice and $\epsilon_{\pm}(\b{k})=\pm t \sqrt{3 + f(\b{k})}$ where 
$f(\b{k})=2 \cos\left( \sqrt{3} k_y \right) + 4 \cos\left( \frac{3 k_x}{2} \right) 
\cos\left( \frac{\sqrt{3} k_y}{2} \right)$~\cite{castro}.
 
\begin{figure}[h!]
\centering
\includegraphics[width=0.80\textwidth]{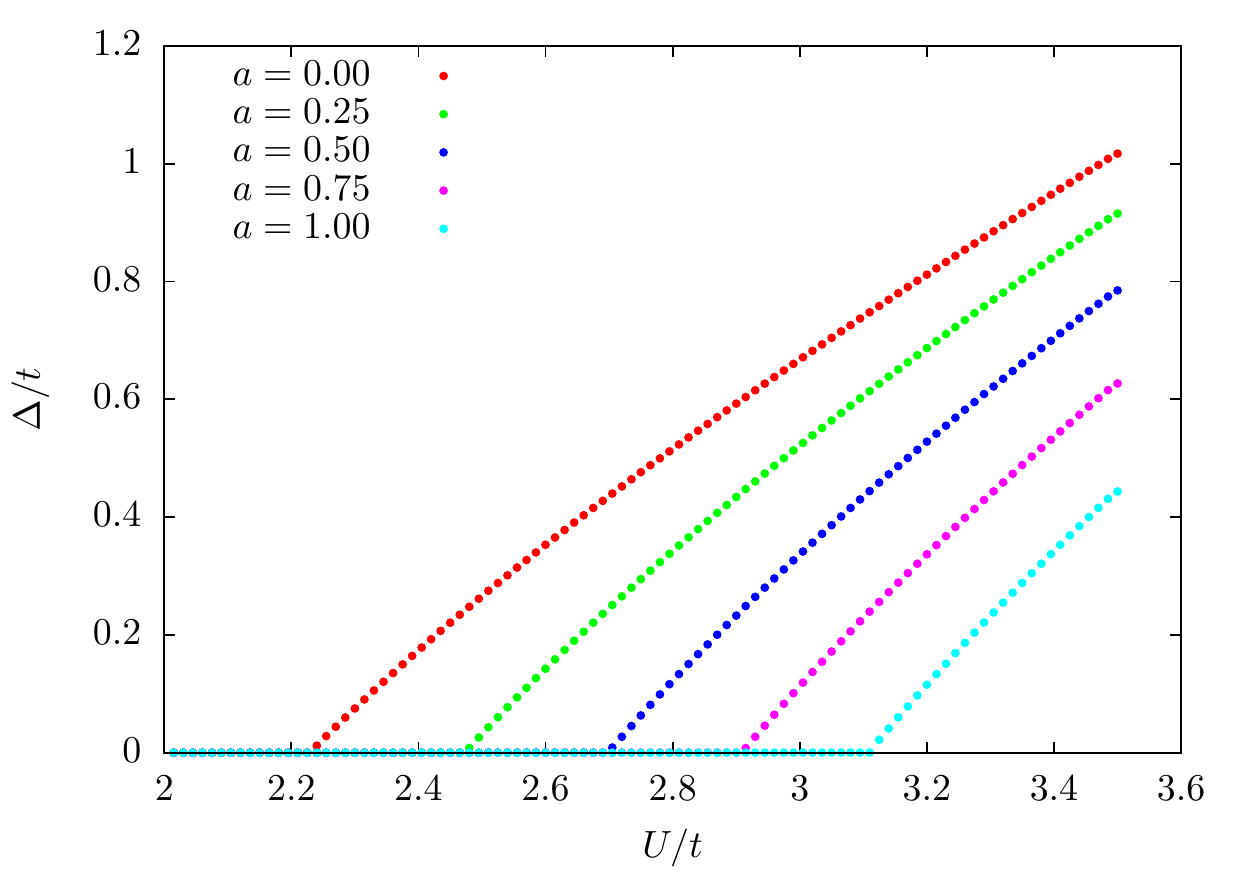}
\caption{Gap $\Delta$ vs. $U$ at $T=0$ for different values of the interpolating parameter $a$ 
and half filling ($n=1$). }
\label{fig:am_honey_gap}
\end{figure}

\begin{figure}[h!]
\centering
\includegraphics[width=0.80\textwidth]{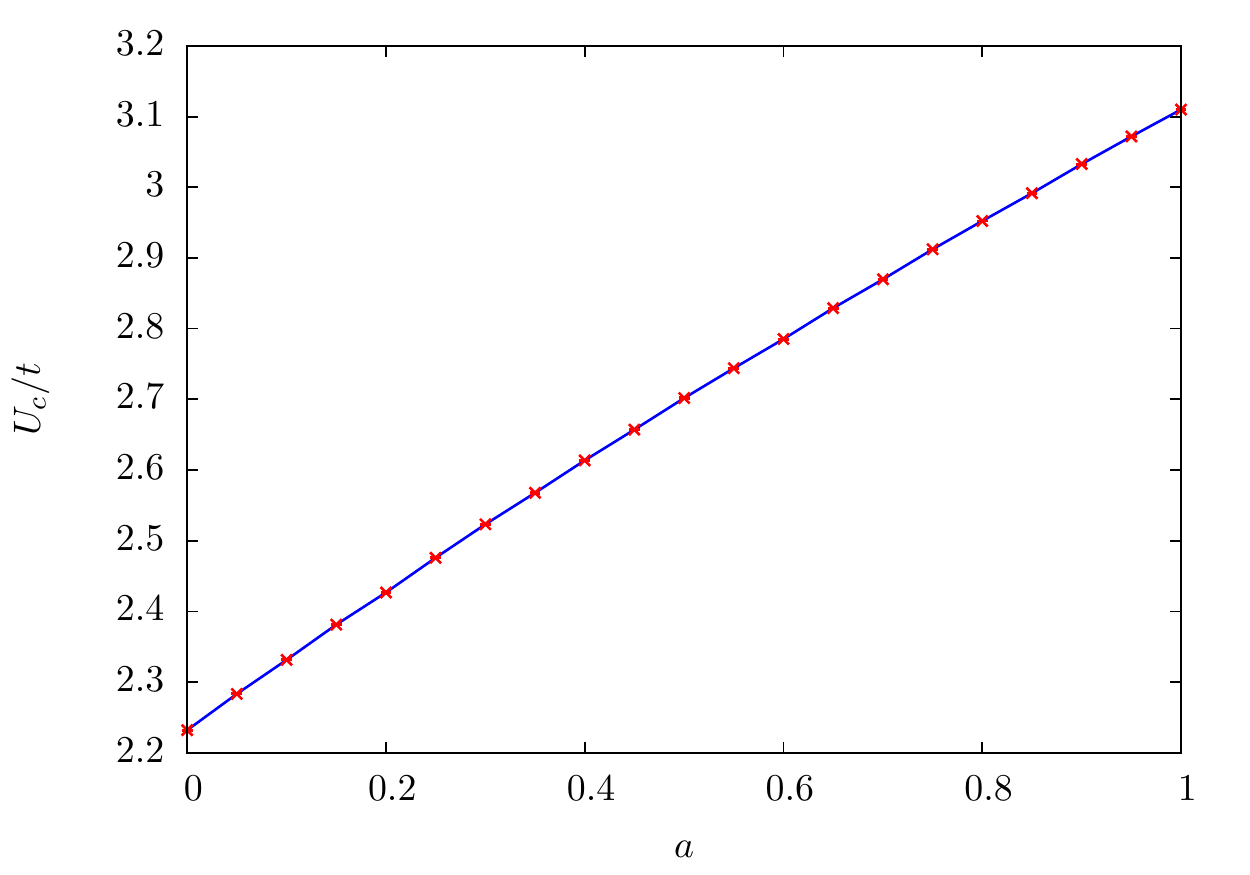}
\caption{Critical value $U_c$ vs. the interpolating parameter $a$ at $T=0$ and half filling (as in Fig. (\ref{fig:am_honey_gap})).}
\label{fig:am_honey_uc}
\end{figure}

\begin{figure}[h!]
\centering
\includegraphics[width=0.80\textwidth]{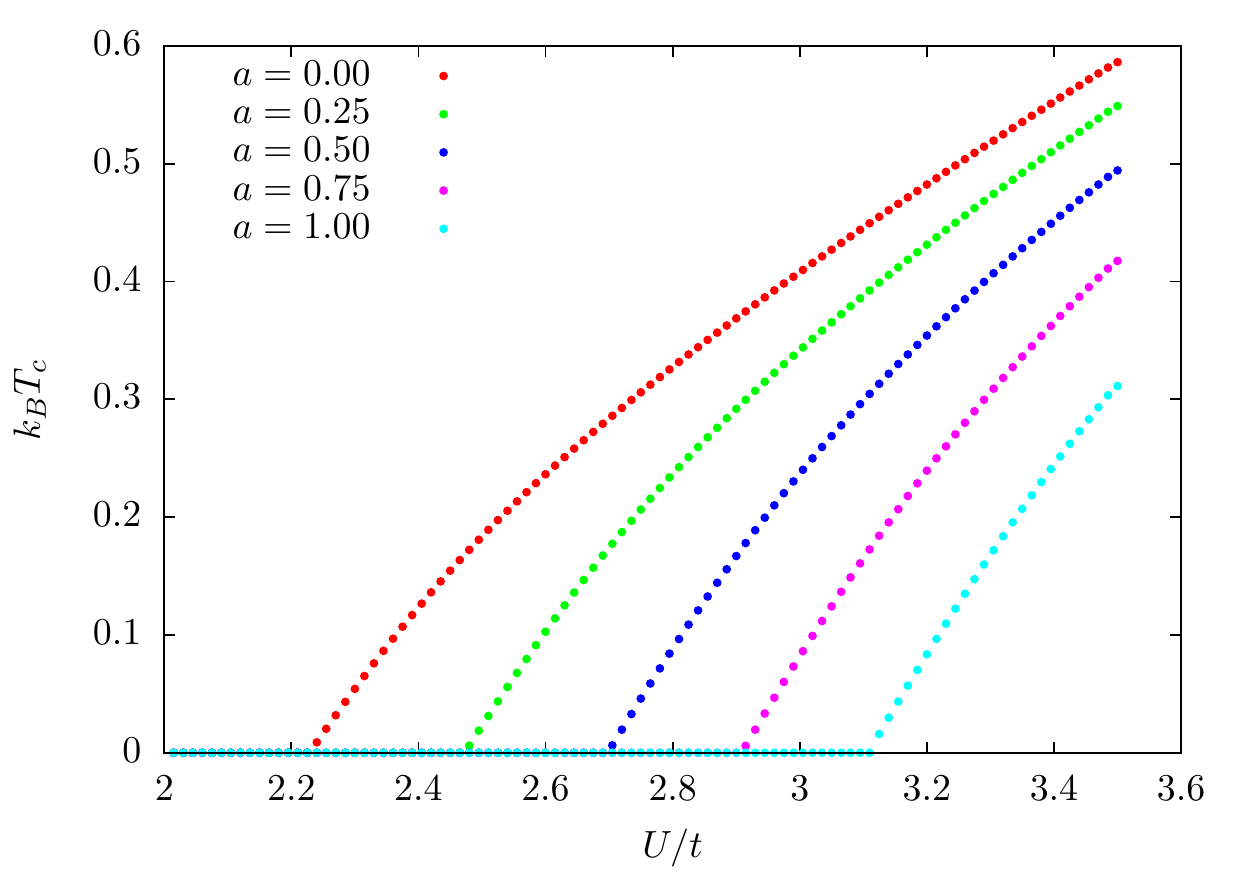}
\caption{Critical temperature $T_c$ (in units of $t$) 
for different values of the interpolating parameter $a$ at half filling (from top to bottom $a=0$, $0.25$, $0.5$, $0.75$, $1$). }
\label{fig:am_honey_tc}
\end{figure}

The mean field results for the gap, the critical temperature and the critical interaction 
(at half filling $n = 1$) are obtained solving eqs. (\ref{mf_eq}) using the single 
particle energy spectrum (\ref{interpol_2D_1}) and (\ref{interpol_2D_2}): the corresponding findings are shown in 
in Figs. (\ref{fig:am_honey_gap}), (\ref{fig:am_honey_uc}) and (\ref{fig:am_honey_tc}). 
In Fig. (\ref{fig:am_honey_gap}) we plot at half filling 
the superfluid gap $\Delta$ vs. $U$ at $T=0$ for different values of the interpolating parameter $a$: 
it is clearly visible in Fig. (\ref{fig:am_honey_gap})  a quantum phase transition 
between a semimetal (for $U$ smaller than a critical value $U_c$) and a superfluid state (for $U>U_c$). The quantum 
phase transition continuously varies passing from the honeycomb lattice to the  $\pi$-flux square and the critical 
value $U_c$ increases with $a$ increasing (i.e., it is larger for the $\pi$-flux square lattice). Results across the interpolation are shown in Fig. (\ref{fig:am_honey_uc}). For the honeycomb lattice ($a=0$) the critical value $U_c$ can be evaluated 
to be $U_c = 2.23 \, t$: this result should be compared with the Monte Carlo calculation 
giving $U_c =(4.5 \pm 0.5) \,  t$~\cite{sorella,mc}. Similarly we expect that the result 
for the critical value of $U_c$ for the $\pi$-flux square lattice is underestimated by the mean field 
(we find $U_c=3.12 \,  t$ for $a=1$). 
Finally in Fig. \ref{fig:am_honey_tc} we plot the mean field 
critical temperature $T_c$ as a function of the interpolating parameter $a$.
Notice that $T_c$ becomes non vanishing at the same critical value $U_c$ for which at $T = 0$ $\Delta$ becomes non vanishing.

A property of the quantum phase transition 
conserved during the interpolation between the honeycomb lattice and the $\pi$-flux square lattice 
is a linear gap scaling law as a function of $|U-U_c|$:
\beq 
\Delta \propto |U -U_c| \, .
\eeq
As it happens in the honeycomb lattice~\cite{sorella}, this is due to the presence 
of $|\Delta|^3$ terms in the Landau-Ginzburg expansion of the Hartree-Fock energy per unit cell at 
zero temperature, which is given for the $\pi$-flux square lattice by the integral 
\beq
\frac{E_{HF} (\Delta)}{\Omega} = \int_{\epsilon_0}^0 \mathrm{d} \epsilon \, g(\epsilon) 
\sqrt{\epsilon^2 + \Delta^2} + \frac{\Delta^2 }{U \Omega},
\label{hf_en_delta}
\eeq
where $\epsilon_0=-2\sqrt{2}t$ is the minimum value of the single particle energy spectrum. Notice that 
(\ref{hf_en_delta}) is also valid for a general value of the interpolating parameter provided 
that $\epsilon_0$ is the minimum energy of the spectrum (\ref{interpol_2D_1}), (\ref{interpol_2D_2}). 
In order to obtain a cubic term in the expansion it is necessary that, at some point within the 
integration interval in eq. (\ref{hf_en_delta}), the density of states is linear 
in $\epsilon$ and $\Delta>\epsilon$ for arbitrary small $\Delta$, then $\epsilon \to 0$. This condition is fulfilled for 
both for the $\pi$-flux square lattice and the honeycomb lattice, as well as across the 
interpolation.

We observe that in presence of a mass term the linear critical exponent of the gap 
is changed: the critical exponent becomes $1/2$, and this again happens across the whole 
interpolation from the honeycomb lattice to the $\pi$-flux square lattice. This can be seen 
by analyzing how the coefficient of the $|\Delta|^3$ terms in the Landau-Ginzburg energy 
changes in presence of 
a mass: we consider for simplicity the honeycomb lattice. Let consider 
an on-site energy $+m$ for the sublattice A and $-m$ for the sublattice B~\cite{gws}: 
the single particle spectrum becomes 
$\epsilon_{\pm}(\b{k})= \pm t \sqrt{3 + f(\b{k})+m^2}$. 
The resulting density of states shows a gap of amplitude $2tm$ around $\epsilon=0$: 
the position of the Van Hove singularities changes introducing the mass term, and the 
density of states diverges for $\epsilon=\pm \sqrt{1+m^2}$. Since 
the density of states is gapped around $\epsilon=0$ so (\ref{hf_en_delta}) 
becomes for the honeycomb lattice:
\begin{align}\label{hf_en_delta_hon}
	\frac{E_{HF} (\Delta)}{\Omega} = \int_{-\epsilon_0(m)}^{-m t} 
\de \epsilon \, g(\epsilon) \sqrt{\epsilon^2 + \Delta^2} 
+ \frac{\Delta^2 }{U \Omega}, 
\end{align}
being $\epsilon_0(m)$ the minimum of the spectrum in the presence of the mass term. 
Therefore, choosing $\Delta>m t$ it is not possible to fulfill the required condition for the cubic term 
in the expansion discussed above,
so that we recover the usual critical exponent ($1/2$).

The previous results for the critical temperature and the gap have been found at mean field level and the inclusion of quantum fluctuations 
on top of it are expected to quantitatively modify such results: in particular in \cite{zhao1} the case of the honeycomb lattice was investigated,
finding that the critical temperature is rather considerably lowered by the inclusion of quantum fluctuations (it is found $T_c^{\mathrm{max}} \sim 0.1 \, t$ \cite{zhao1}). 
Proceeding as in \cite{zhao1}, we performed for $a=1$ an estimate of the Kosterlitz-Thouless critical temperature $T_{\mathrm{KT}}$: we found a decrease of the critical temperature induced by quantum fluctuations less pronounced then for the honeycomb at $ a= 0$ (e.g. for $a = 1$ and  $U \sim 1.5 \, U_c $ we found that $T_{\mathrm{KT}} \sim 0.5  \, T_c$, where $T_c$ is the mean field critical temperature). Given the smoothness exhibited by the considered model across the interpolating path $a = [0,1]$, we expect the presented results for the gap and the critical temperature will be quantitatively modified of a similar amount (but qualitatively robust) passing from $a = 0$ to $a=1$. A qualitative difference is anyway expected to be induced by the quantum fluctuations for the specific values of the critical exponents for the gap near $T_c$. More importantly time-dependent fluctuations around mean field coupling amplitude and phase fluctuations are required to study the collective modes, as the Leggett and the Goldstone modes.

\section{3D lattice with Dirac points: $\pi$-flux cubic lattice model}
\label{3sect}

The discussion presented in the previous Section shows that at mean field level the superfluid 
properties of the attractive Hubbard model on the honeycomb lattice and on the 
$\pi$-flux square lattice are qualitatively equivalent, with only quantitative differences. 
In order to study 
the corresponding problem in $3D$ it is then natural consider the cubic $\pi$-flux lattice, which has 
the desired structure of the Dirac points (we remind that stacking up layers of graphene 
leads to the graphite geometrical structure, with Dirac cones destroyed~\cite{Wallace}). 
We consider a cubic lattice with magnetic field $\b{B}$ along the direction $\b{d} = (1,1,1)$ 
such to generate a $\pi$-flux on each plaquette. This can be obtained using the Peierls 
substitution with the vector potentials~\cite{Has} $$\b{A}(\b{r}) = \pi (0, x-y, y-x).$$ 

Notably enough, tuning the hopping amplitude along the $z$ direction it is possible to investigate a 
crossover between $2D$ and $3D$ lattice having the same Dirac points: this can be done considering 
hopping amplitudes with absolute value $t$ along $\hat{x}$ and $\hat{y}$ direction and $t_a\equiv a t$ along $\hat{z}$. 
$a$ is an interpolation parameter ($0 \leq a \leq 1$) characterizing how much connected are the layers of the system: 
changing the value of the parameter $a$ one can explore the bidimensional $\pi$-flux square 
lattice model ($a=0$) and the isotropic $3D$ $\pi$-flux square lattice model ($a=1$). 

\begin{figure}[h!]
\centering
\includegraphics[width=0.65\textwidth]{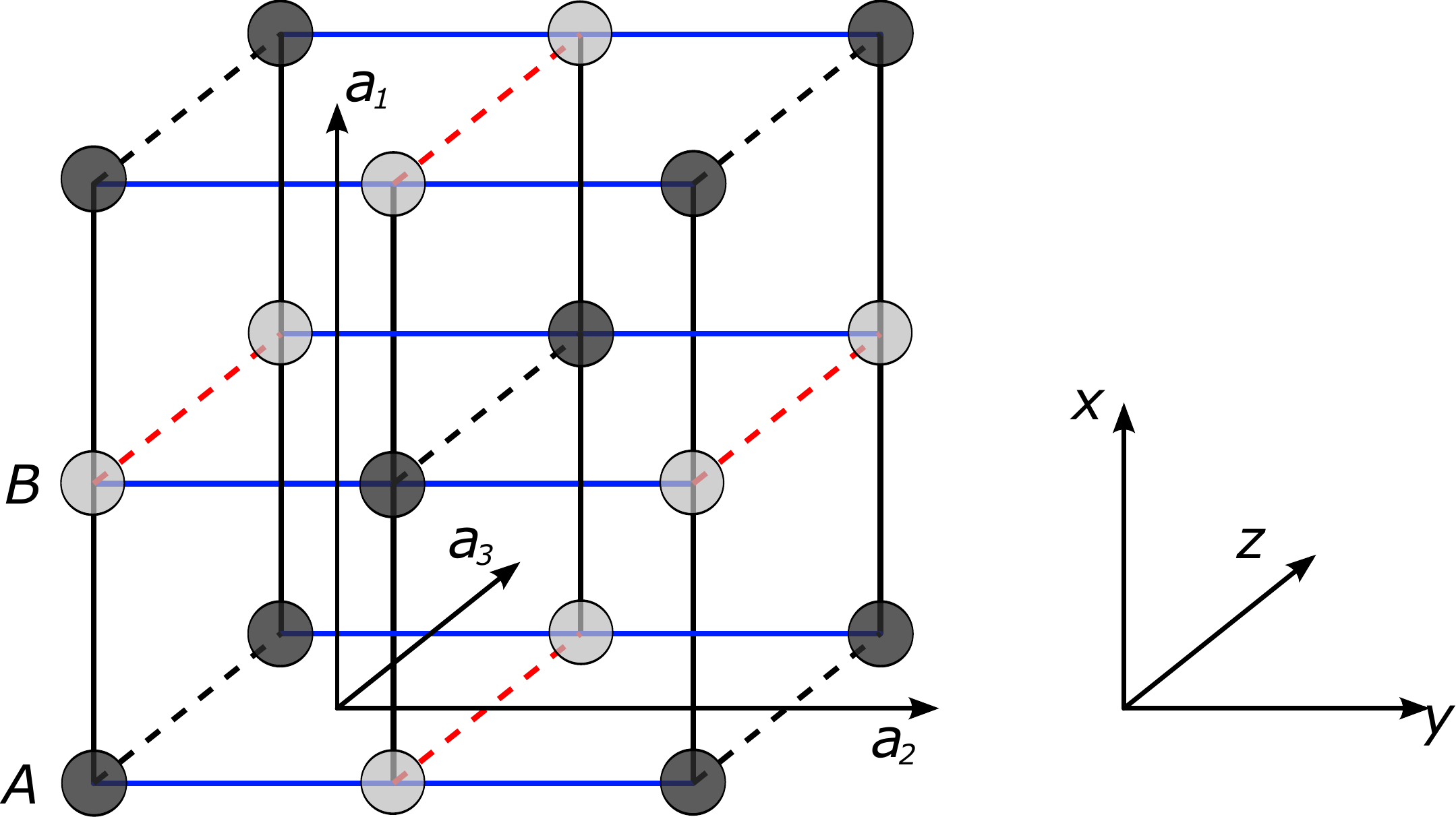}%
\caption{Layered $\pi$-flux cubic lattice model - different phases of the 
hopping amplitudes are represented with different colors: 
$0 \, \mod \, (2\pi)$ (black bonds along the $x$-direction and bonds along 
the $z$-direction joining darker sites), $\pi \,  \mod \, (2\pi)$ (red bonds along 
the $z$-direction joining clearer sites) and 
$\pi /2  \, \mod \, (2\pi)$ (blue bonds in the $y$-direction). 
Primitive lattice vectors are also shown.}
\label{fig:am_3d_lattice}
\end{figure}

\begin{figure}[h!]
\centering
\includegraphics[width=0.65\textwidth]{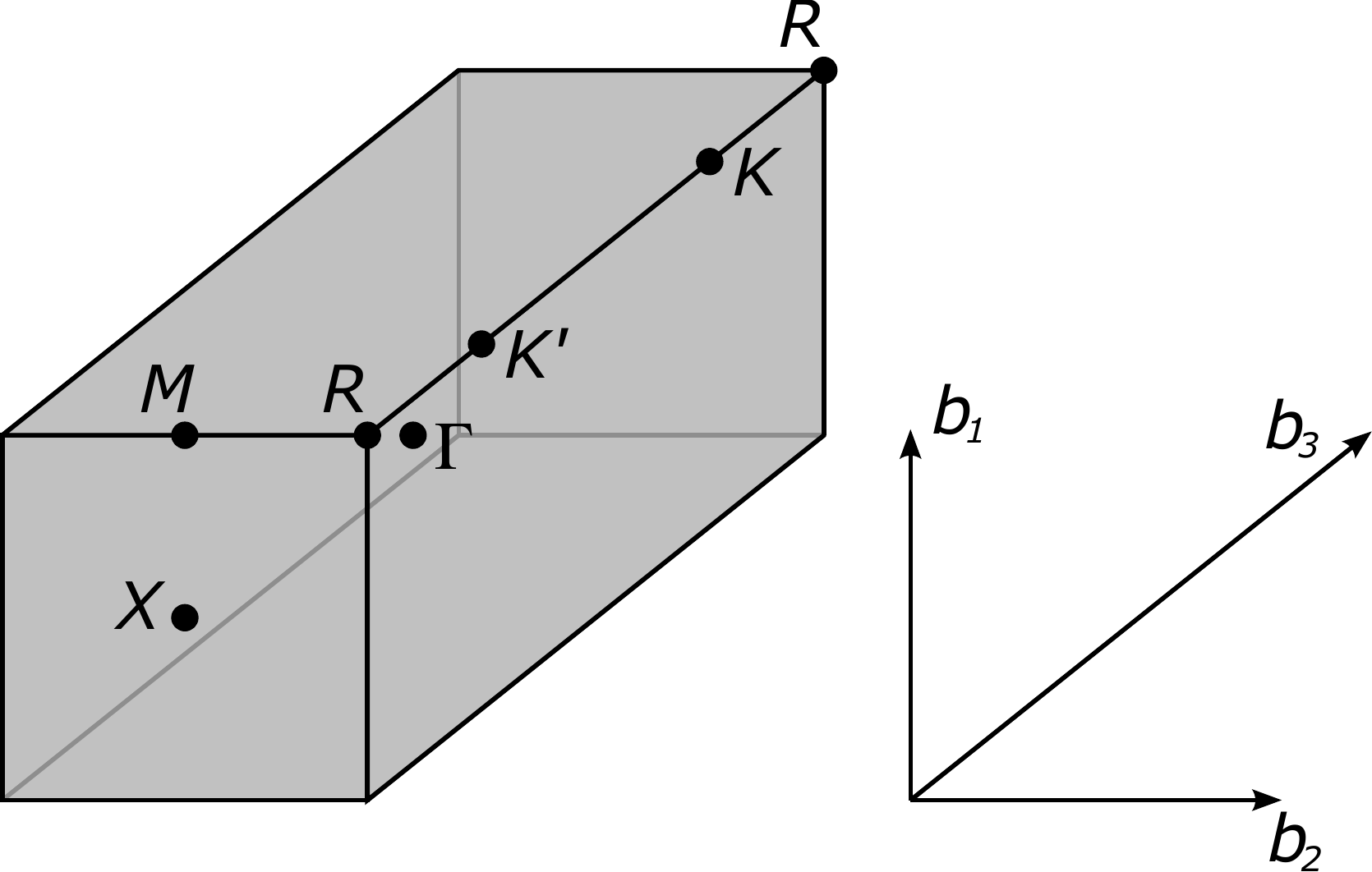}%
\caption{First Brillouin zone for the layered $\pi$-flux cubic lattice model. 
Reciprocal lattice vectors are also shown.}
\label{fig:am_3d_bz}
\end{figure}

Fig. (\ref{fig:am_3d_lattice}) draws a schematic plot of the layered $\pi$-flux cubic lattice. 
The lattice vectors in units of the lattice spacing (assumed to be equal in all the three directions) 
are given by
\begin{align}
	\b{a}_1=\left(2 ,0,0 \right) \qquad \b{a}_2= \left( 0 ,2,0  \right) \qquad  \b{a}_3= \left( 0,0,1  \right) 
\end{align}
so the reciprocal lattice vectors are
\begin{align}
	\b{b}_1= \pi \left( 1 ,0 ,0 \right) \qquad \b{b}_2=\pi \left(0, 1 ,0 \right) \qquad \b{b}_3=2\pi \left(0, 0 ,1 \right).
\end{align}
Notice that the primitive cell is four times the one relative to the cubic lattice with no magnetic flux. 
This is due to the presence of $\b{B}$: the translations along $\hat{x}$, $\hat{y}$, $\hat{z}$ 
do not commute each others any longer and one has to double two of them. 
This doubling results in halving two sides of the first Brillouin zone cell, as plotted in 
Fig. (\ref{fig:am_3d_bz}).

\begin{figure}[h!]
\centering
\includegraphics[width=0.80\textwidth]{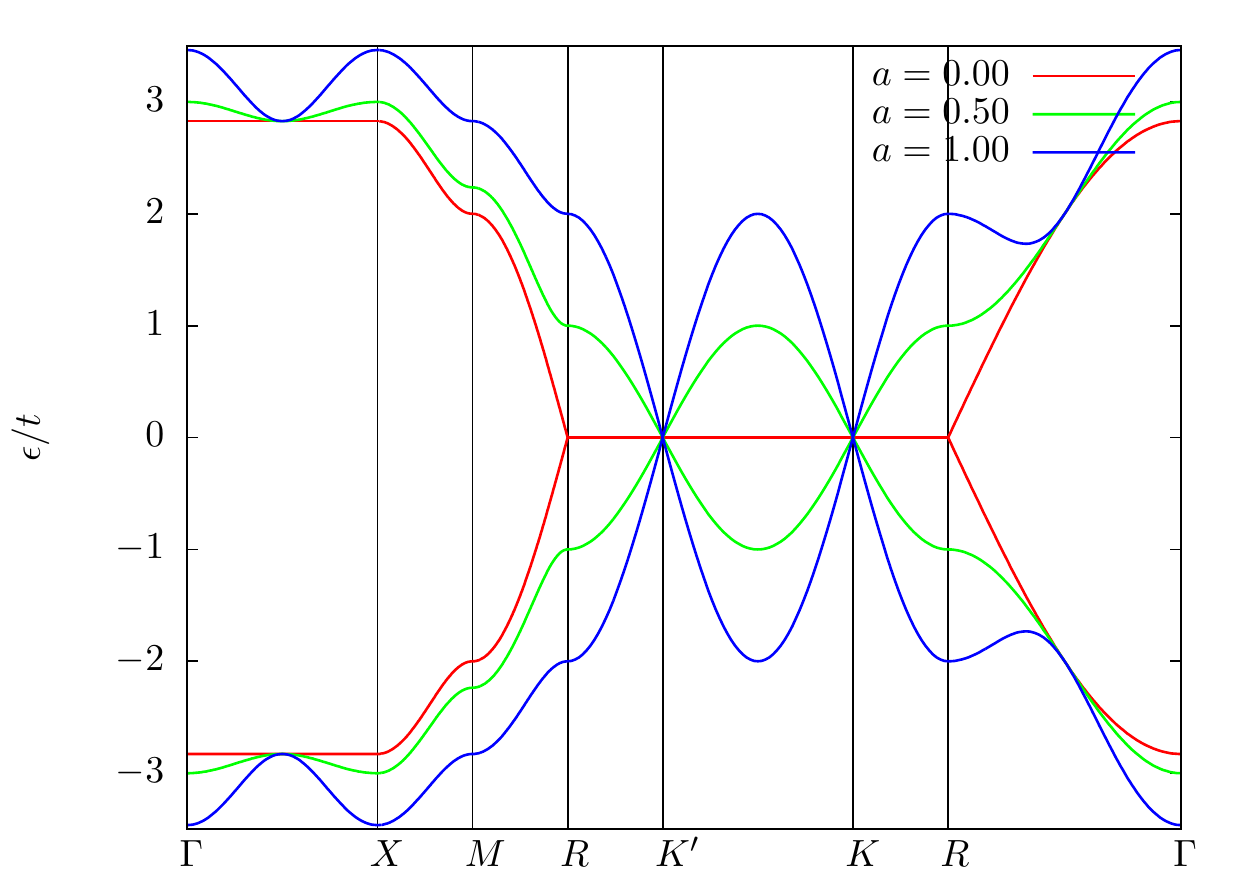}%
\caption{Dispersion relation for the Hamiltonian $\op{H}_0$ 
of the $\pi$-flux cubic lattice model for different values of the parameter $a$ 
(from top to bottom $a=0$, $0.5$, $1$).}
\label{fig:am_3d_band}
\end{figure}

\begin{figure}[h!]
\centering
\includegraphics[width=0.80\textwidth]{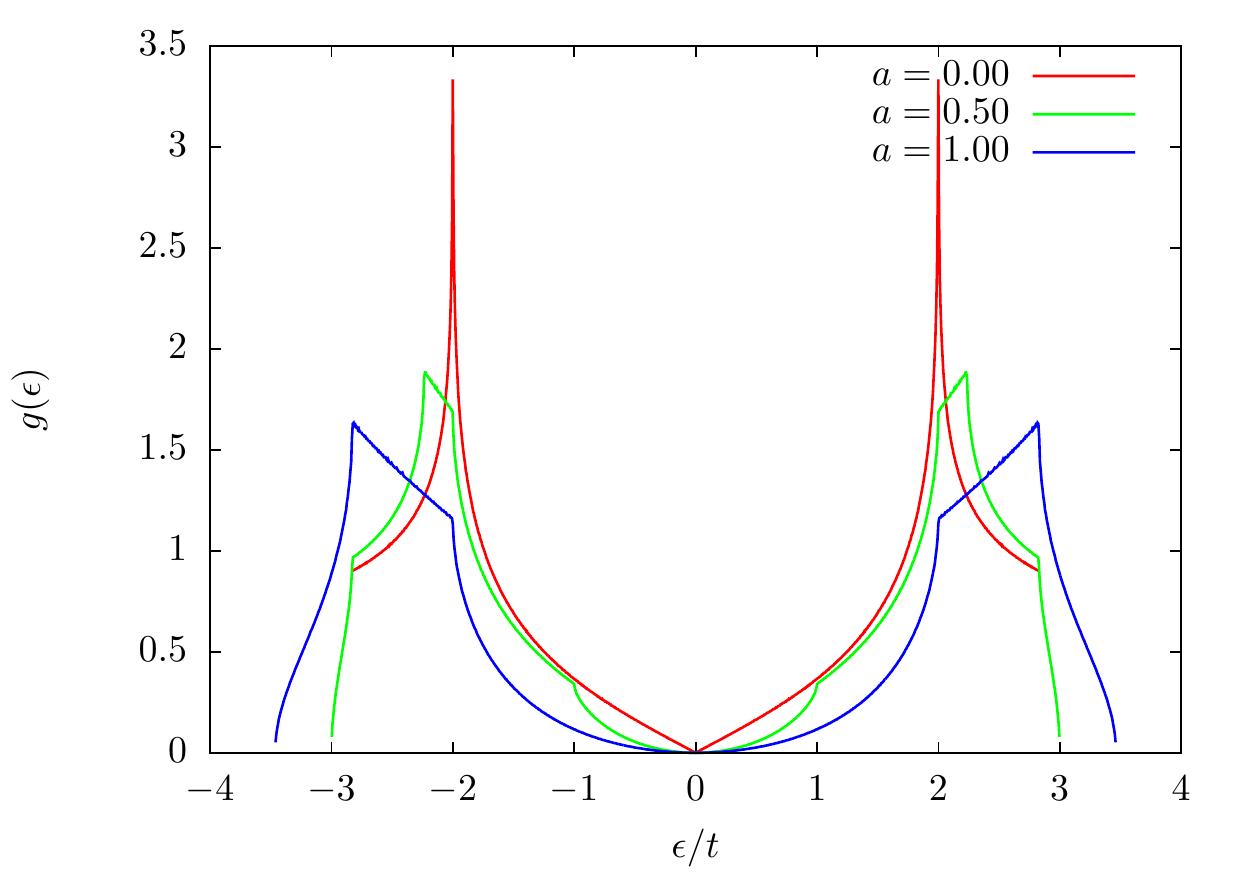}%
\caption{Density of states for the Hamiltonian on the $\pi$-flux cubic 
lattice model for different values of the anisotropy parameter $a$ 
(from top to bottom $a=0$, $0.5$, $1$).}
\label{fig:am_3d_dos}
\end{figure}

The Hamiltonian $\op{H}_0$ in the Fourier space is given by
\begin{align}\label{am_nn_3D}
	\op{H}_0=2t \sum_{\b{k}\in \mathrm{1^{st}BZ},\sigma} \begin{pmatrix} \cop{c}_{\b{k} \sigma,a} \cop{c}_{\b{k} \sigma,b}\end{pmatrix}   \begin{pmatrix} a \cos (k_z+\pi) & 2 \cos k_x + 2 \mathrm{e}^{-i \frac{\pi}{2}} \cos k_y\\2 \cos k_x + 2 \mathrm{e}^{i \frac{\pi}{2}} \cos k_y  & a \cos k_z \end{pmatrix}\begin{pmatrix} \aop{c}_{\b{k} \sigma,a} \\ \aop{c}_{\b{k} \sigma,b}\end{pmatrix}
\end{align}
where the subscript $a$ and $b$ indicates the $respectively \, A$ and $B$ sublattices shown in Fig. (\ref{fig:am_3d_lattice}).
This Hamiltonian can be diagonalized exactly, leading to the single particle energy spectrum
\begin{align}
	&\epsilon_{\pm}(\b{k}) = \pm \, 2 t \, \sqrt{\cos ^2 k_x + \cos^2 k_y + a^2 \cos ^2 k_z }.
\label{spectrum3D}
\end{align}
The energy spectrum has two inequivalent Dirac points 
\begin{align}
	\b{K}=\frac{\pi}{2} \left( 1 ,1, 1 \right) \quad \mbox{and} \quad \b{K}^{\prime}=\frac{\pi}{2} \left( 1 ,1,-1 \right) \, , 
\end{align}
notably their position does not depend on the layering parameter $a$. 
The dispersion relation around these points is given by
\begin{align}
	\epsilon_{\pm}(\b{K}+\b{q}) = \pm 2 t |\b{q}| \sqrt{1+ (a^2-1)\cos^2 \theta} + \mathcal{O}\left( |\b{q}|^3 \right).
\end{align}
The linear order term depends on the polar angle $\theta = \arcsin \left( q_z / |\b{q}| \right)$, 
so the speed of light for the Dirac Fermions is not isotropic (unless $a=1$). 
Of course, if $a=0$ this quantity vanishes along the $z$ direction. 
The band structure and the density of states for different values of the parameter $a$ are shown 
in Figs. (\ref{fig:am_3d_band}) and (\ref{fig:am_3d_dos}) respectively. 
It is also possible to obtain an expansion for the density of states near $\epsilon=0$ if $0<a \le 1$:
\begin{align}
	g(\epsilon) \approx \frac{\epsilon^2}{2 a \pi^2 t^3} \, .
\end{align}

\begin{figure}[h!]
\centering
\includegraphics[width=0.80\textwidth]{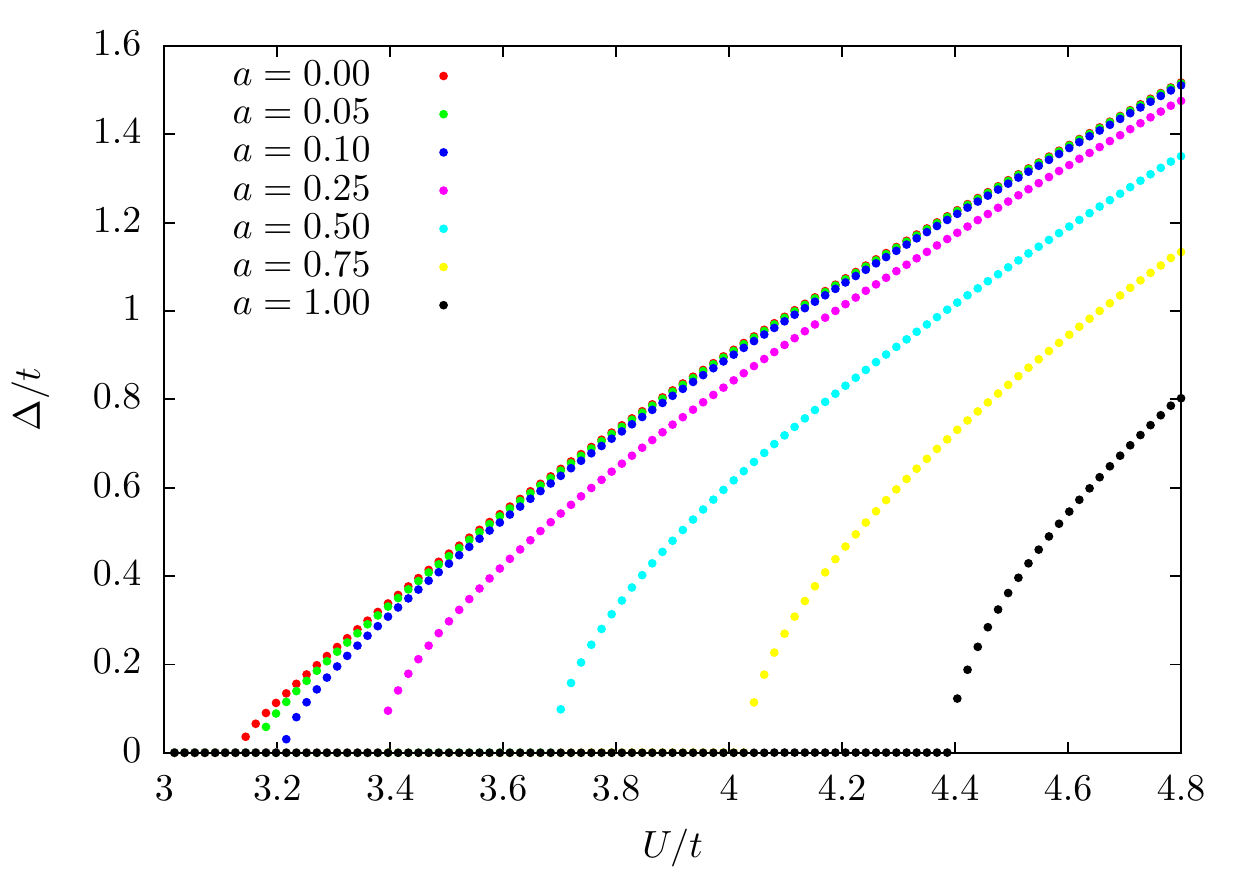}
\caption{Gap $\Delta$ vs. $U$ for different values of the anisotropy parameter $a$ 
for the $\pi$-flux cubic lattice model at half filling (from top to bottom is $a=0$, $0.05$, $0.1$, $0.25$, $0.5$, $0.75$, $1$).}
\label{fig:am_honey_gap_3D}
\end{figure}

\begin{figure}[t!]
\centering
\includegraphics[width=0.80\textwidth]{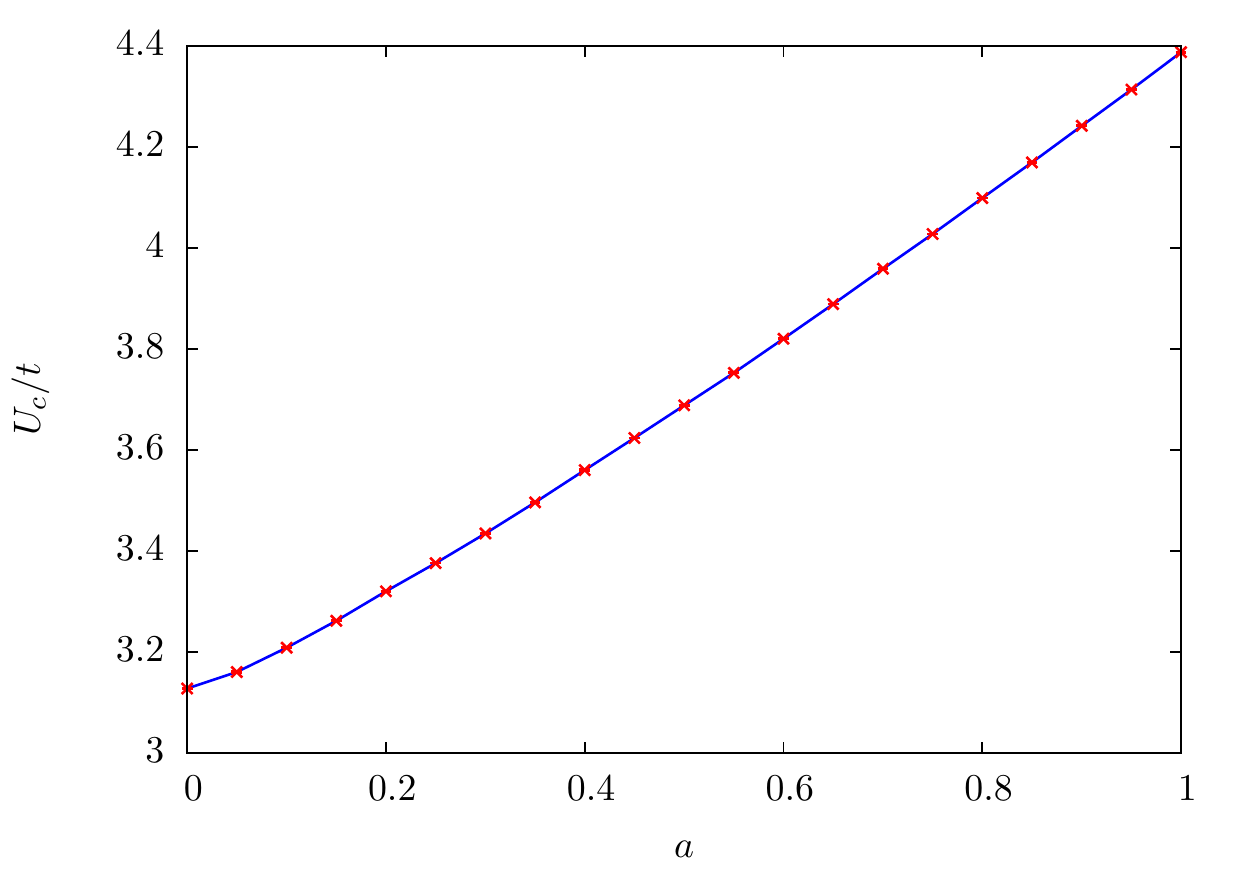}
\caption{Critical interaction $U_c$ for different values of the anisotropy parameter $a$ for 
the $\pi$-flux cubic lattice model at half filling.}
\label{fig:am_honey_uc_3D}
\end{figure}

\begin{figure}[h!]
\centering
\includegraphics[width=0.80\textwidth]{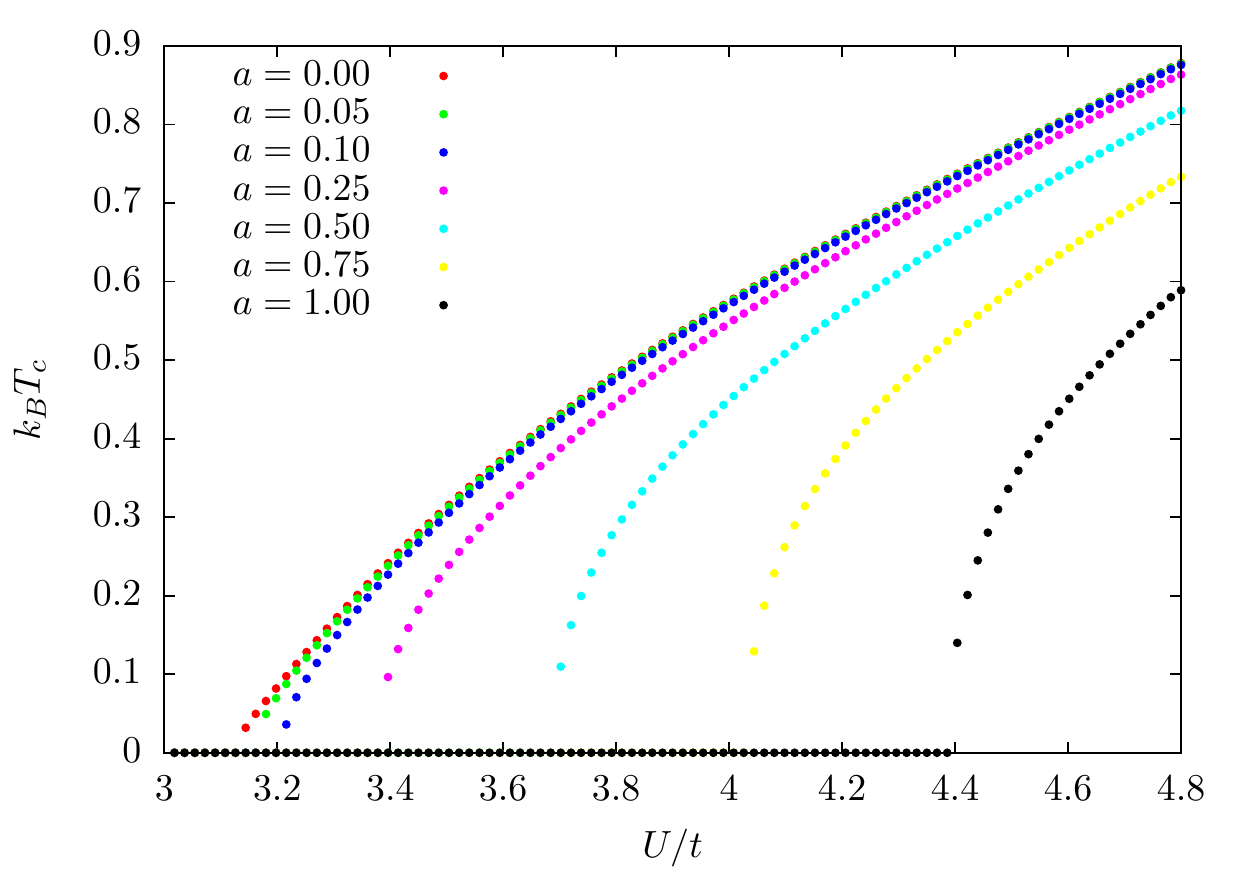}
\caption{Critical temperature $T_c$ for different values of the anisotropy parameter $a$ for 
the $\pi$-flux cubic lattice model at half filling (from top to bottom is $a=0$, $0.05$, $0.1$, $0.25$, $0.5$, $0.75$, $1$).}
\label{fig:am_honey_tc_3D}
\end{figure}

We solved the BCS mean field equations at half filling with the single particle energy spectrum 
(\ref{spectrum3D}): the various results are shown in 
Figs. (\ref{fig:am_honey_gap_3D}),
(\ref{fig:am_honey_uc_3D}) and (\ref{fig:am_honey_tc_3D}).
The quantum phase transition between semimetal and superconductor is robust passing from a 
$2D$ to a $3D$ system, as shown in Fig. (\ref{fig:am_honey_gap_3D}): at mean field 
level the change is quantitative, a larger interaction being needed to induce 
the superfluid phase as shown in Fig. (\ref{fig:am_honey_uc_3D}). Consequently, mean field critical 
temperature decreases with $a$ increasing, as shown in 
Fig. (\ref{fig:am_honey_tc_3D}).

We  observe that the critical exponent $\alpha$ in the law 
$\Delta \propto |U -U_c|^{\alpha}$ at $T=0$ abruptly changes from $1$ to the usual value $1/2$ for finite $a$ 
because of the different behavior of the density of states near to the Fermi level. 
In fact, introducing the hopping terms along the $z$ direction, $g(\epsilon)$ 
is not linear in the neighbourhood of the Fermi energy, as shown in Fig. \ref{fig:am_3d_dos}. 
We finally point out that for a quantitative improvement of the results presented in this Section the inclusion of quantum fluctuations is certainly needed, even though we expect on general grounds that such corrections are less pronounced compared to the 2D cases discussed in Section \ref{2sect}. For this reason we think that an interesting line of future work would be to include quantum fluctuations passing from 2D to 3D, i.e. from $a=0$ to $a=1$, and possibly to include the effect of charge-density waves on equilibrium properties at half filling.  

\section{Conclusions}

In this paper we analyzed the superfluidity of attractively interacting fermions 
in a family of $2D$ and $3D$ ultracold lattice systems with Dirac points, focusing in particular 
on the honeycomb lattice, the $\pi$-flux square lattice, the $\pi$-flux cubic lattice and 
related interpolating geometries.
We studied the effects of interaction and dimensionality on the relevant physical parameters, 
as the superfluid gap, chemical potential and critical temperature $T_c$: 
these quantities are found to vary continuously along the patterns of interpolation.  
In the $2D$ cases, at zero temperature and at half filling there is a quantum phase transition occurring at 
a critical (negative) interaction $U_c$ presenting a linear critical exponent for the gap as a function of $|U-U_c|$. 
This behavior holds for the honeycomb lattice, for the $\pi$-flux square lattice model and for the 
interpolating schemes. We also observed that across the interpolation between the honeycomb and the 
$\pi$-flux square lattice, in presence of an energy offset for the two sublattices (i.e., a mass term) 
the critical exponent is the usual one ($1/2$). These investigations show that at mean field level there 
is no qualitative difference between the superfluid properties of the attractive honeycomb model 
and the $\pi$-flux square lattice.  Following \cite{zhao1}, we also performed an estimate of the effect of quantum fluctuations on the mean field critical temperature, finding that the decrease of $T_c$ is less pronounced on the $\pi$-flux square lattice with the respect of the honeycomb lattice.   

In three dimensions the quantum phase transition is again retrieved and it is shown to be continuously varying 
passing from the $\pi$-flux square lattice to the $\pi$-flux cubic lattice:  we point out that, unlike the $2D$ case, the critical exponent for the gap changes from $1$ to $1/2$.

We observe that in the two considered interpolations (from the honeycomb to the $\pi$-flux square lattice and from the $\pi$-flux cubic lattice to the $\pi$-flux square lattice) a smooth behavior for the gap and the critical temperature is found: since the considered lattices all have Dirac points, but different spectra far from them, our obtained results show that most of the contribution to the physical observables come - at mean field level - from low energy excitations around Dirac points.
 
Our analysis was done at mean field level: since the effects of non-mean field terms are in general 
relevant - especially for the critical temperature~\cite{zwerger12} - further investigations for the considered lattices 
are needed to include the effects of fluctuations, whose effects have been studied in honeycomb lattices 
at half filling~\cite{zhao1} and estimated for the $\pi$-flux square lattice in this paper. Similarly, it would be very interesting to study the effects on the superfluidity 
of a general truly non-Abelian synthetic gauge potential for the $3D$ $\pi$-flux cubic lattice: 
work on this direction is currently on-going.

\section*{Acknowledgements}
Discussions with D. Giuliano, M. Capone, A. Maraga, M. Burrello and M. Iazzi are very gratefully acknowledged. 
L.L. acknowledges a grant awarded by Banco de Santander and financial support from 
European Regional Development Fund.

\end{document}